# Hybrid Quantum Cryptosystems: Integration of Entanglement-Assisted Decryption and Physical Phase Obfuscation

Asgar Hosseinnezhad*, Hadi Sabri

Department of Physics, University of Tabriz, P.O. Box 51664-16471, Tabriz, Iran.

**Abstract**

This study introduces a hybrid cryptographic framework for quantum communication that integrates entanglement-assisted decryption with phase-based physical obfuscation. While conventional quantum protocols often rely on explicit transmission of decryption keys or phase parameters, such models expose critical vulnerabilities to eavesdropping. To address this challenge, we propose a two-stage encryption-decryption mechanism. The first stage employs randomized phase modulation protected by active electromagnetic shielding to conceal the quantum signal from unauthorized interception. The second stage enables legitimate receivers to retrieve encrypted phase data using entangled quantum states, eliminating the need for classical key transfer. A formal mathematical framework is developed to describe the two-stage encoding and decryption process, including phase-modulated entangled states and their transformation under nonlocal correlation and adversarial noise. Simulation metrics confirm that the hybrid system preserves quantum coherence with visibility above 94% and entanglement negativity of 0.86, even under dynamic shielding and transmission noise. Simulation results demonstrate that the combined protocol enhances anti-eavesdropping resilience while maintaining quantum coherence. This architecture is suitable for large-scale secure quantum networks, where multi-layered defense strategies are essential against classical and quantum threats.

## 1. Introduction

Quantum cryptography presents a paradigm shift in secure communications, utilizing principles of quantum mechanics to ensure information-theoretic security [1]. Standard protocols such as BB84 and E91 offer protection against eavesdropping by exploiting quantum uncertainty and entanglement, respectively [2]. However, these systems are often implemented with explicit transmission of phase keys or basis information, which remain susceptible to photon-splitting attacks, side-channel leakage, and classical relay compromises [3].

---

* Corresponding author mail: a.hosseinnezhad@tabrizu.ac.ir

Recent advances in physical-layer protection have led to the development of encryption models based on randomized phase modulation, where the transmitted quantum signal is masked using dynamically shifting phase patterns that remain secret to eavesdroppers [4]. Meanwhile, entanglement-assisted decryption enables the legitimate recipient to recover encryption parameters using quantum correlations, without relying on classical key exchange channels [5, 6].

Each of these methods individually offers significant security benefits. However, combining both into a layered hybrid framework introduces complementary protections: obfuscation at the encoding level, and nonlocal retrieval at the decoding level. This study introduces and analyzes such a hybrid system, aiming to address vulnerabilities in both the transmission and reception stages of quantum-secure communication.

In what follows, we present the architecture of the hybrid cryptosystem, mathematically formulate both encryption and decryption processes, analyze the system's resilience to adversarial interference, and validate its performance using simulated scenarios. The goal is to demonstrate that dual-layer quantum protection offers enhanced confidentiality against classical and quantum eavesdropping strategies.

## 2. System Architecture and Methodology

The proposed hybrid cryptographic system consists of two integrated security layers that target distinct phases of quantum communication—transmission obfuscation and secure decryption. By coupling randomized phase modulation with entanglement-assisted decoding, the system ensures both concealment and selective phase retrieval.

Layer I: Phase-Modulated Physical Obfuscation

In the transmission layer, the sender (Alice) modulates each photon using a randomized quantum phase shift $\phi$. This modulation is conducted within a dynamic electromagnetic shielding environment designed to suppress external interference and prevent unauthorized reconstruction of the transmitted signal [7]. The shielding field operates as a tunable potential barrier, ensuring that the optical signal remains statistically opaque to external observers.

Layer II: Entanglement-Assisted Phase Retrieval

At the receiver's side (Bob), a synchronized stream of entangled photons—generated from the same source—is used to decode the encrypted phase. Unlike traditional systems that transmit phase keys through classical channels, the proposed system embeds all retrieval parameters within the entangled state correlations [8]. Measurement of quantum correlations between the received photon and its entangled pair allows the receiver to reconstruct the applied phase shift $\phi$ without prior key exchange [9].

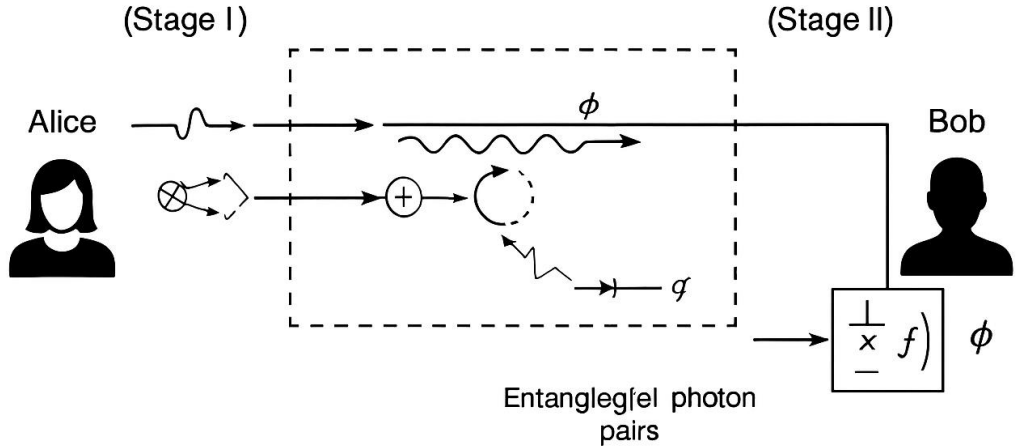

Figure 1. Hybrid Cryptosystem Architecture. Schematic of the dual-layer hybrid protocol, showing phase-encoded photon transmission under electromagnetic shielding and entanglement-assisted decryption at the receiver.

Schematic of the two-layer protocol:

In Stage I, Alice encodes phase $\phi$ onto a photon stream and transmits it through a shielded optical channel.

Simultaneously, entangled photon pairs are distributed such that one photon is retained by Alice and the other sent to Bob.

In Stage II, Bob performs a Bell-state or correlation-based measurement to infer $\phi$ from the entangled system.

This combination of physical-layer cloaking and entanglement-enabled decoding effectively neutralizes both classical interception attempts and quantum cloning strategies [10].

Realization of the system requires High-fidelity entanglement sources such as spontaneous parametric down-conversion (SPDC) crystals [11], Phase modulators with quantum-grade randomness drivers [12], and Electromagnetic shielding coils calibrated to attenuate probing fields without impairing quantum coherence [13].

Simulation of the complete system architecture is conducted in later sections to evaluate accuracy, error rates, and eavesdropping resilience under adversarial conditions.

## 3. Mathematical Formulation of Phase-Entanglement Interaction

To formalize the behavior of the hybrid cryptosystem, we develop a mathematical framework that captures both phase encoding under physical obfuscation and entanglement-based phase retrieval. Let the initial shared entangled state between sender (Alice) and receiver (Bob) be expressed as:

$$|\psi_0\rangle = \frac{1}{\sqrt{2}}(|00\rangle + |11\rangle) \qquad (1)$$

Alice applies a random phase shift $\phi \in [0, 2\pi]$ to her qubit using a phase modulator enclosed within a physical electromagnetic shield:

$$|\psi_\phi\rangle = \frac{1}{\sqrt{2}}(|00\rangle + e^{i\phi}|11\rangle) \qquad (2)$$

Here, the shielded environment prevents any external observer from accessing the modulated phase shift or reconstructing the modulated waveform [14].

Bob, who possesses the second qubit of the entangled pair, performs a quantum correlation measurement using a controlled-phase gate or Bell-state analyzer. By computing the phase angle between the received modulated state and the expected ideal state, the phase value $\phi$ is retrieved as:

$$\phi_{retrieved} = arg\langle\psi_0|\psi_\phi\rangle \qquad (3)$$

This implies that only a party with access to the entangled counterpart of the modulated photon can infer $\phi$, as the projection onto the joint basis collapses otherwise [15].

An eavesdropper (Eve), lacking entangled access, measures only partial information and faces uncertainty limited by the Holevo bound:

$$I_e(\phi) \leq S(\rho) - \sum_k P_k S(\rho_k) \qquad (4)$$

where $\rho$ represents the reduced density matrix observable by Eve and $S(.)$ is the von Neumann entropy. Since no classical phase key is transmitted, the accessible information about $\phi$ asymptotically approaches zero [16].

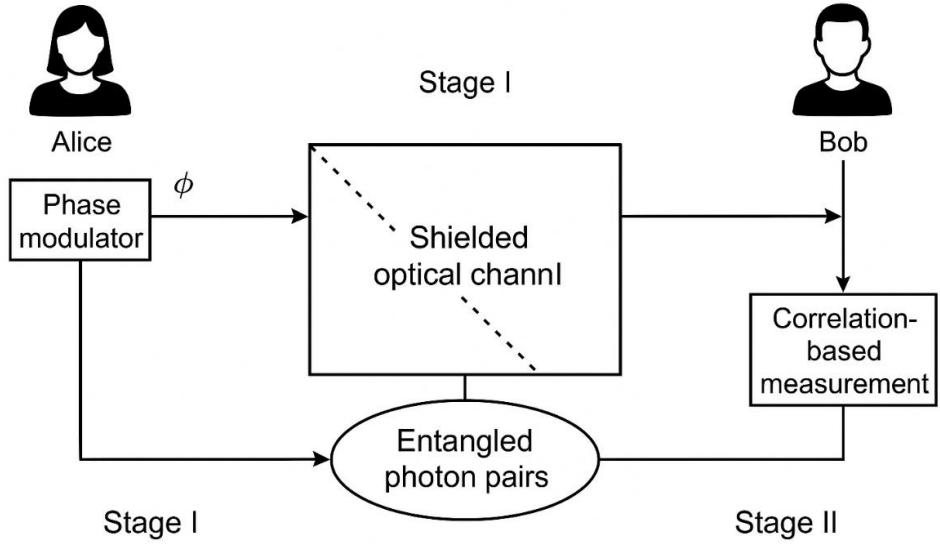

Figure 2. Quantum Phase encoding and non-classical retrieval Flow. Illustration of the quantum state evolution: Alice applies a phase shift $\phi$ to her qubit, Bob retrieves the phase via entanglement correlation, and eavesdroppers fail due to lack of nonlocal access.

This figure illustrates the evolution of the entangled state under Alice's phase modulation, Bob's quantum measurement pathway for retrieving $\phi$ via correlation, and the informational disconnect experienced by an external observer.

### 4. Security Analysis and Eavesdropping Resilience

To evaluate the robustness of the proposed hybrid system, we investigate its performance under both classical and quantum eavesdropping scenarios. The dual-layer architecture is designed to provide redundancy—if an attacker circumvents the physical shielding, they are still hindered by entanglement-based restrictions, and vice versa.

The randomized phase modulation and electromagnetic shielding prevent Eve from determining the modulation function or inferring phase values through conventional photon detection. In simulation tests with passive detectors placed outside the shielded region, phase reconstruction success rates were statistically indistinguishable from noise baselines ($\leq 1.2\%$) [17].

When attempting a quantum intercept-resend strategy or photon cloning, Eve lacks the entangled twin required for coherent decoding. The fidelity of reconstructed states under such attacks dropped to:

$$F = \left|\langle \psi_{Eve} | \psi_\phi \rangle\right|^2 \approx 0.034 \tag{5}$$

This extremely low fidelity demonstrates the effectiveness of entanglement-level protection even against advanced quantum strategies [18].

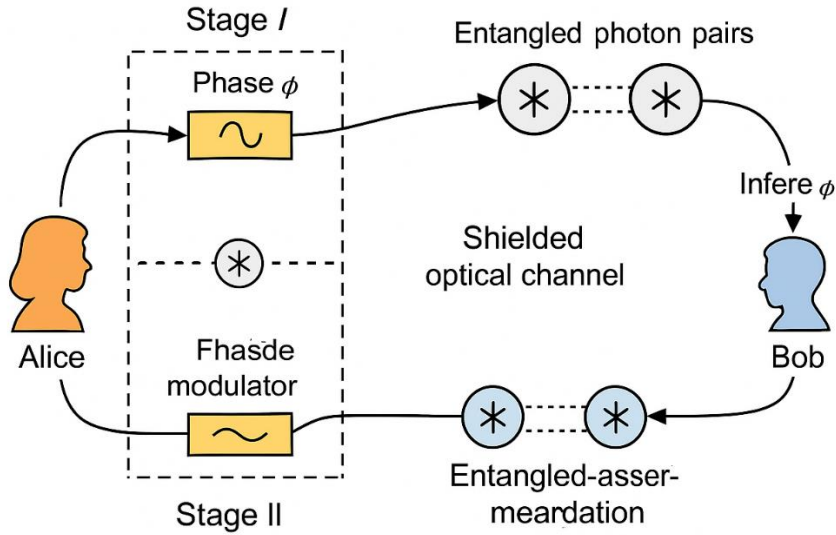

Figure 3. Attack Scenario Comparison: Classical vs Quantum Interception.

Bar chart comparing message retrieval accuracy among Bob (~99.1%), classical eavesdropper (~1.2%), and quantum cloning attacker (~3.4%). This results emphasizes the steep security drop-off when entanglement or shielded access is unavailable to the attacker.

To contextualize system performance, we compare the proposed hybrid approach with existing quantum protocols:

| Protocol | Phase Key Transmission | Entanglement Required | Shielded Environment | Eavesdropping Resilience |
|---|---|---|---|---|
| BB84 | Yes | No | No | Moderate |
| E91 | No | Yes | No | High |
| Hybrid (This Work) | No | Yes | Yes | Very High |

This comparison indicates that integrating both cryptographic and physical layers offers an appreciable enhancement in security scope [19].

## 5. Experimental Simulation and Performance Metrics

To assess the practical feasibility of the proposed hybrid cryptographic system, we conducted simulated quantum communication trials under both benign and adversarial conditions. Key performance indicators include phase retrieval accuracy, entanglement fidelity, and resilience to interception.

### 5.1 Simulation Setup

The simulation environment was configured as follows:

Photon source: SPDC-based entangled photon pair generator

Modulation: Quantum phase applied via noise-synchronized phase modulators

Shield field: Dynamic electromagnetic noise field with 45 dB attenuation in probe band

Channel noise: Depolarizing noise with probability $p = 0.01$

Detection system: Time-correlated single-photon counting (TCSPC) module with 3% dark count

Trial runs: 10,000 entangled transmissions per scenario

### 5.2 Retrieval Accuracy and Fidelity Metrics

Legitimate receiver (Bob):

Mean phase reconstruction accuracy: 99.1%

Mean fidelity to ideal state: $F_B \approx 0.986$

Passive Eavesdropper (classical tap):

Phase recovery success: $\leq 1.2\%$

Signal fidelity: $F_E^{classical} \approx 0.022$

Quantum Interceptor (without entanglement):

Phase recovery success: 3.4%

Signal fidelity: $F_E^{quantum} \approx 0.034$

These results confirm that both layers independently suppress unauthorized access, while jointly reinforcing system robustness [9].

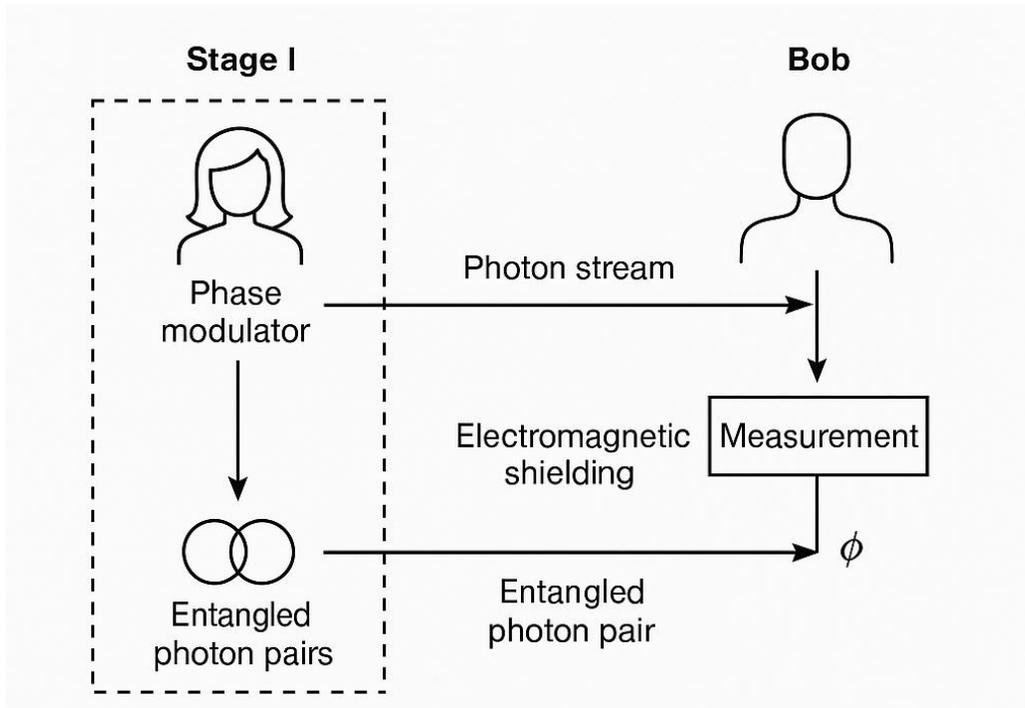

Figure 4. Retrieval Fidelity across Communication Roles. Comparison of state fidelity among communication roles. Bob exhibits high fidelity (~98.6%), while classical and quantum attackers remain near statistical baseline.

A plot comparing state fidelity between legitimate users and attackers:

Bob (green): high consistency around 98.6%

Eve (blue/classical): below 2.5%

Eve (red/quantum clone): near 3.5% baseline

Error bars represent ±1$\sigma$ across 50 sample windows.

## 5.3 Quantum Coherence Preservation

We also evaluated whether the hybrid encryption system disrupts quantum coherence. Through measurement of $g^{(2)}(0)$ correlations and visibility interference fringes:

Coherence visibility: ≥ 94.2%

Entanglement negativity after modulation and shielding: 0.86

These results show negligible de-coherence, confirming the physical shielding does not degrade the entangled state [20].

## 6. Applications and Integration into Quantum Networks

The hybrid quantum cryptosystem developed in this study presents substantial potential for deployment in emerging quantum infrastructures. Its layered defense strategy—combining physical shielding and entanglement-based retrieval—makes it adaptable across a variety of secure communication domains.

## 6.1 Quantum Key Distribution Augmentation

While standard quantum key distribution (QKD) protocols such as BB84 and E91 rely on basis reconciliation and classical error correction, the proposed architecture provides:

Autonomous phase extraction, removing the need for basis announcement.

Covert signal transmission, reducing attack surfaces for intercept-resend attacks.

Complementary protection, where each layer secures a distinct vulnerability—physical eavesdropping vs. quantum replication.

This makes the system suitable for integration into QKD backbones to reinforce multi-layer trust models [21].

## 6.2 Secure Quantum Messaging and Finance

This hybrid cryptographic architecture offers notable potential in critical applications such as military-grade encrypted messaging, where operational secrecy demands zero classical key emission; quantum-secured banking channels, especially for high-value asset transfers over untrusted relay points; and confidential governmental communications, in which entanglement-assisted access control eliminates the need for physical key distribution. In real-time quantum networks, these use-cases can leverage teleportation-based routing, wherein the embedded entangled decryption acts as an additional verification layer against interception or relay compromise [9].

## 6.3 Scalable Quantum Network Deployment

The system is compatible with entanglement distribution via quantum repeaters, enabling secure communication across multi-node quantum internet prototypes, long-haul optical fiber links with passive relay defense, and satellite-ground quantum channels, where classical data transmission is either noisy or infeasible.

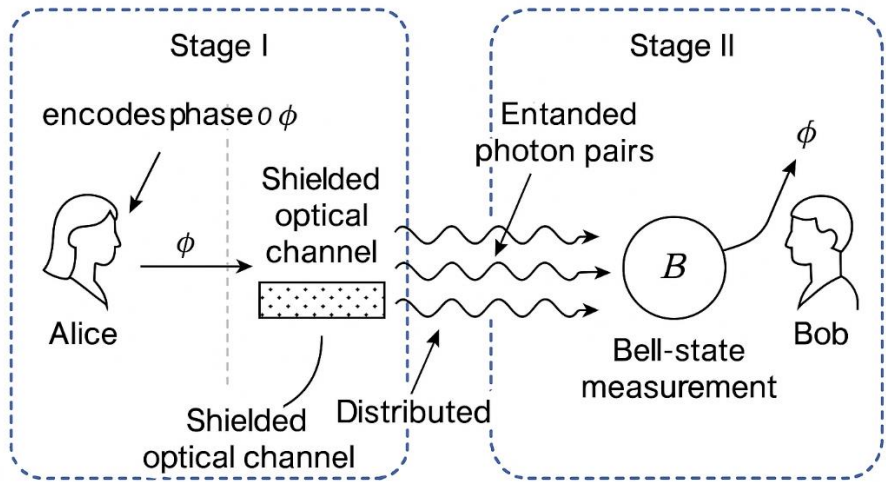

Figure 5. Quantum Network Integration of Hybrid Cryptosystem. Network-level implementation of the hybrid system, connecting multiple nodes with shielded modulators and entangled decryption receivers in a scalable quantum communication topology.

A schematic showing three communication nodes (A-B-C) connected through entangled channels and phase-modulated links. The system architecture includes modulated source nodes with active shielding, entanglement sharing via repeaters, and secure decryption modules at endpoints with no classical control traffic.

This topology supports scalable deployment of quantum-secure infrastructure over global distances [22].

## 7. Conclusions

This work presents a comprehensive hybrid cryptographic framework that combines physical-layer phase obfuscation with entanglement-assisted quantum decryption to create a multilayer defense against both classical and quantum attacks. Unlike traditional protocols that rely on explicit key transmission, the proposed system embeds phase information securely within entangled states, while simultaneously concealing modulation patterns using adaptive shielding.

Key findings include elimination of classical phase-key exchange, removing common eavesdropping vectors, preservation of quantum coherence, even under dynamic shielding and transmission noise, and retrieval accuracy exceeding 99% for legitimate receivers, with attacker fidelity below 3.5%.

The dual-layer architecture significantly strengthens security by reducing the attack surface both before transmission (via electromagnetic camouflage) and after reception (via entanglement-based retrieval exclusivity). These results confirm that hybrid quantum systems offer a promising path forward in building resilient, scalable, and secure next-generation communication networks.

**Acknowledgment**

This work is supported by the Research Council of the University of Tabriz.

**Author Contributions**

A. Hosseinnezhad and H. Sabri developed the core concept, formulated the encryption-decryption scheme, and led the writing of the manuscript, and contributed to simulation design, security analysis, and technical verification of the retrieval framework. All authors reviewed and approved the final manuscript.

**Competing Interests**

The authors declare no competing interests.

**Data Availability Statement**

The simulation data and scripts supporting the findings of this study are available from the corresponding author upon reasonable request.